\documentclass[titlepage,12pt]{article}
\usepackage[utf8]{inputenc} 
\usepackage[T1]{fontenc}    
\usepackage{hyperref}       
\usepackage{url}            
\hypersetup{
  colorlinks   = true, 
  urlcolor     = blue, 
  linkcolor    = blue, 
  citecolor   = blue 
}

\usepackage{subcaption}     
\usepackage{graphicx}      

\usepackage{tabularx}       
    \newcolumntype{L}{>{\raggedright\arraybackslash}X}
\usepackage{longtable}    
\usepackage[table]{xcolor}
\usepackage{booktabs}       
\usepackage{amsfonts}       
\usepackage{nicefrac}       
\usepackage{microtype}      
\usepackage{amsmath}
\usepackage{mathtools}
\usepackage{amssymb}
\usepackage{float}
\usepackage[margin=1 in]{geometry}
\title{A Comparison of Functional Principal Component Analysis Methods with Accelerometry Applications}
  \author{Bohan Wu\\
    Department of Statistics, Rice University\\
    and \\
    Bradley Van Allen \\
    Department of Statistics, Rice University\\
    \\
    Advisor: Dr. Daniel Kowal\\
    Department of Statistics, Rice Univerity\\
    }
\date{May 2021}
\begin{document}
\maketitle

\section*{Abstract}
 The association between a person's physical activity and various health outcomes is an area of active research. The National Health and Nutrition Examination Survey  (NHANES) data provide a valuable resource for studying these associations. NHANES accelerometry data has been used by many to measure individuals' activity levels. A common approach for analyzing accelerometry data is functional principal component analysis  (FPCA). The first part of the paper uses Poisson FPCA  (PFPCA), Gaussian FPCA  (GFPCA), and nonnegative and regularized function decomposition  (NARFD) to extract features from the count-valued NHANES accelerometry data.  The second part of the paper compares logistic regression, random forests, and AdaBoost models based on GFPCA, NARFD, or PFPCA scores in the context of mortality prediction. The results show that Poisson FPCA is the best FPCA model for the inference of accelerometry data, and the AdaBoost model based on Poisson FPCA scores gives the best mortality prediction results.\\
\section{Introduction}
The association between a person's daily physical activity levels and various health outcomes is an active research area. The National Health and Nutrition Examination Survey (NHANES) is a large annual survey conducted by the Centers for Disease Control (CDC)  to collect health, behavioral and nutritional information of the US population (CDC, 2017).  NHANES is one of the most important national health surveys in the world in terms of size, scope and availability. Furthermore, NHANES was the first study to make public a large dataset containing the physical activity information measured using accelerometers when they released accelerometry data from the 2003 - 2004 and 2005 - 2006 participants (Leroux et al., 2019).  \\
\\
The accelerometer is  a hip-worn device that detects and records the magnitude of acceleration or “intensity” of movement according to a specified time interval. NHANES collected its 2003 - 2004 and 2005 - 2006 participants' physical activity data with integer-valued "activity count" measured by the accelerometers (CDC, 2017), and processed raw measurements into minute-by-minute activity data.   In this study, our focus is on the NHANES accelerometry data from 2005 - 2006 samples. We are interested in  extracting features from NHANES accelerometry data and building a prediction model using the extracted accelerometry features along with the demographic and comorbidity covariates from NHANES. \\
\\
A common feature extraction approach for accelerometry and wearable data is functional principal component analysis (FPCA). Functional data are defined as data consisting of random functions, where each function is considered as a sample from a stochastic process (Shang, 2014). Functional Data Analysis  (FDA) is an emerging field that studies statistical methodologies that analyze functional data, most notably by Ramsay and Silverman (1997). FDA is able to extract the information shared across smooth functions, and it is most often applied to high-dimensional time series that assume a latent continuous data generative process. One of the most widely studied methods in FDA is functional principal component analysis  (FPCA). FPCA is an adaptation of principal component analysis (PCA) to the analysis of function-valued observations. Several previous studies have used FPCA to extract the accelerometry features {(Crainiceanu et al., 2009, Leroux et al., 2019)}. The theoretical properties and computation of FPCA models have been studied in great detail  ({Shang, 2014}). However, most applications of FPCA have focused on analyzing Gaussian functional data, with relatively little literature on other types of functional data. In recent years, however, there have been rising interests in modeling count-valued functional data in wearable technology ({Backenroth et al., 2020}), ecology {(Müller et al.,  2005)}, and finance {(Bening and Korolev, 2012)}.  \\
\\
For NHANES accelerometry data, count-valued functional methods would be advantageous for inference. The traditional approach is to transform the count data and apply the Gaussian FPCA { (Leroux et al., 2019)}. However, these transformations are not well-defined for integer-valued data-generative process, and log transformation has difficulty dealing with zero-inflation { (O'Hara et al., 2010)}. As a remedy, generalized FPCA extends the ideas from Gaussian FPCA to distributions from the exponential family ({Goldsmith, 2017}). The generalized FPCA family includes Poisson FPCA  (PFPCA), which is conceptually more well-defined for modeling functional count data by capturing some aspects of the integer-specific properties through a Poisson generative process. Besides FPCA approaches, another promising FDA method is nonnegative and regularized function decomposition (NARFD). NARFD is a close alternative to PFPCA that provides interpretable, local parameter estimates and reconstruction for nonnegative count-valued functional data ({Backenroth et al., 2020}).\\
\\
Our first application of FPCA methods involves using GFPCA, PFPCA, and NARFD to extract features from and conduct inference on the accelerometry data of 2005 - 2006 NHANES cohort. We provide detailed interpretations for the functional components and scores from GFPCA, PFPCA, and NARFD, and determine the most suitable model for interpreting the participants' physical activity. The second application involves using the functional scores from the three FPCA methods to predict the five-year all-cause mortality outcomes of NHANES 2005 - 2006 participants in the year 2011 which was made available by linking the 2005 - 2006 NHANES cohort to the year 2011 death certificates {(Leroux et al., 2019)}. The mortality outcome is defined as whether or not a given participant had died from any cause within five years of the original survey. We combine GFPCA, PFPCA, and NARFD scores with typical covariates in NHANES as predictors in logistic regression, random forest, and AdaBoost methods to predict mortality outcomes, and we discuss the strength and weakness of each method in terms of predictive performance. Finally, we determine the variable contribution to mortality and provide insights relevant to the medical community. 

\section{Accelerometry Feature Extraction}
We use Gaussian FPCA, Poisson FPCA, and NARFD to extract functional principal component (FPC) scores from the accelerometry data. The FPC scores serve as the summary measure of the participant's physical activity. This allows for the study of associations between individual physical activity,  health outcomes, and future mortality. Furthermore,  we compare the interpretability of the three methods to determine the most interpretable method for the accelerometry data. In particular, we judge the model interpretability based on the goodness of fit quantified with MAE, interpretability of the FPCs, model assumptions, and how well the fitted values capture the data's smooth, functional patterns. 
 
\subsection{Data}
The National Health and Nutrition Examination Survey (NHANES) is a large, stratified, multistage, annualized survey conducted by the Centers for Disease Control (CDC) that collects health and nutrition data on the non-institutionalized US population. The NHANES website affirms that the data are collected with the intent that it will be analyzed to “help develop sound public health policy, direct and design health programs and services, and expand the health knowledge for the Nation” (CDC, 2017). NHANES is one of the largest and most influential public health surveys in terms of size, scope, diversity, and public availability of the data. Furthermore, NHANES was the first study to publicize a dataset that contains the physical activity information measured with accelerometers when they released accelerometry data from the 2003 - 2004 and 2005 - 2006 participants. The accelerometry data were collected on the participants selected for the Mobile Examination Center (MEC) portion of the study. For this study, we analyze the accelerometry data from the 2005 - 2006 NHANES cohort. 
\subsubsection{Accelerometry Data}
The accelerometry data consists of 5-minute level accelerometry data reported as "activity counts" for NHANES 2005 - 2006 participants. It is a dataset with 6269 rows for 1011 subjects and 288 columns that represent 5-min time points. The raw data come from the package \textit{rnhanesdata} {(Leroux et al., 2019)} that contains minute level accelerometry data reported as "activity counts" for NHANES 2005-2006 participants. The accelerometry data are recorded each minute and contains minute $0$ to minute $1440$ in a typical day. A related dataset, also from the \textit{rnhanesdata} package, contains a flag denoting whether or not the accelerometer was worn at each minute of the day. To tackle missing values, we first re-code activity counts which are considered "non-wear" to be 0. This does not impact many data points as most estimated non-wear times already correspond to 0 counts. To reduce the anomalies and missing values, we aggregate the minute-level data into 5-min bins using summation, which leaves $288$ time points. Then, we remove all $5$-min aggregated curves with missing values. \\
\\
Although each participant may have up to seven days of accelerometry records in the raw data, we summarize the accelerometry records of each subject with one record consisting of the rounded median activity account for the subject at each 5-minute time point. Furthermore, we only include the participants with no missing values in all NHANES covariate categories, with more details in Section \ref{subsect:adata}. The processed data contains $1011$ rows and $288$ columns. The rounded median summary pairs each participant with one unique accelerometry record and allows us to implement Poisson-type methods by preserving the count-valued structure. For accelerometry feature extraction, all $1011$ subjects are included.
\subsection{Methods}
\label{subsect:accel}
Extracting features from accelerometry data is challenging for three reasons. The high dimensionality of the accelerometry data makes it difficult to apply the usual dimensionality reduction techniques such as PCA.  Moreover, accelerometry data assumes a continuous data generative process that requires the method to produce smooth fitted values. Finally, NHANES accelerometry data are count-valued by design with properties such as zero-inflation and over-and-under-dispersion. The ideal model needs to capture these count-specific properties. To address those issues, we investigate three methods: 1) GFPCA with log transformation, 2) PFPCA, and 3) NARFD. \\
 \subsubsection{Gaussian Functional Principal Component Analysis}
Let $Y_i (t)$ be a real-valued response curves for $t \in \tau$ with mean $\mu (t)$ and covariance function $\sigma (t, t')$. In practice, the nonegative count data follow a discrete time grid $Y_i (t_j)$ for subject $i = 1,..., N$ at time $j= 1,..., T_i$. The Gaussian functional principal component analysis  (GFPCA) model has the following generative process: 
\begin{equation}
      Y_i (t) = \mu (t) + \sum_{k = 1}^K s_{ik}\phi_k (t) + \epsilon_i (t)  \quad \epsilon_i (t) \sim N (0, \sigma^2),
\end{equation}
where $s_{ik} \sim N (0,\lambda_k)$, and $(\lambda_k$ and $\phi_k (t))$ are eigenvalue/eigenfunction of the covariance function with $\lambda_1 \geq \lambda_2 \geq \dots  \geq 0$. The set of functions $\{\phi_j (t) \}_{j = 1}^K$ is orthonormal and can be modelled as a linear combination of basis functions such as splines. We implement a log transformation on the responses by taking $log(Y + 1)$ for each response curve $Y$. \\
\\
The Gaussian FPCA is fast and easy to implement. We implemented the Gaussian FPCA via the "fpca.face" function in the \textit{refund} package (Goldsmith, 2020). However, the Gaussian method is not well-defined for count-valued data: the data generative process of Gaussian FPCA cannot reproduce discrete responses, which amplifies model misspecification, limits intepretability and harms the reliability of the inference and predictive distributions ({Kowal and Canale, 2020}). For this reason, we would like to move beyond GFPCA and explore methods that are well-defined for count-valued data. 
\subsubsection{Poisson Functional Principal Component Analysis}
The Poisson functional principal component analysis  (PFPCA) with a logarithmic link function has the following generative process: 
\begin{equation}
    Y_i (t) \sim Pois[exp (\mu_i (t))],\quad \mu_i (t) = \mu (t) + \sum_{k = 1}^K s_{ik}\phi_k (t) + \epsilon_i (t), 
\end{equation} 
where the parameter definitions are identical to those of GFPCA. The model parameters of Poisson FPCA are estimated using a mixed model-based two-step approach  (Goldsmith et al., 2017). Implementation of Poisson FPCA and other generalized FPCA methods in R is available in the \textit{registr} package (Wrobel et al., 2020). \\
\\
Although Poisson FPCA has slower runtime than Gaussian FPCA, the Poisson model is well defined for count-valued data, and the log link mitigates the Poisson distribution's disadvantage to model zeros. Poisson FPCA has all the inference and prediction tools that Gaussian FPCA has, with less chance for model misspecification and more reliable inference. 
\subsubsection{Nonnegative and Regularized Function Decomposition}
The nonnegative and regularized function decomposition (NARFD) method is a novel decomposition of nonnegative functional count data that draws on the concept of nonnegative matrix factorization ({Backenroth et al., 2020}). Compared to FPCA methods, NARFD uses a local, instead of global, estimation scheme. Assume that the observations are Poisson-distributed. NARFD uses the following generative process for functional observations: 
\begin{equation}
    Y_i (t) \sim Pois\{\mu_i (t)\}, \quad \mu_i (t) = \sum_{k = 1}^K s_{ik} \phi_k (t), 
\end{equation}
where $\phi_k (t), k \in 1,..., K$ are functional prototypes, similar to functional principal components (FPCs) in FPCA, and $s_{ik}$ are scores. The estimated function $\mu_i (t)$ is a nonnegative latent process characterizing both the mean and variance of $Y_i (t)$. Expanding the functional prototypes in terms of spline basis functions with $M$ dimensions, we can express the $\mu_i (t)$ term as a discrete-time process:
\begin{align*}
    \mu_i (t_j) = \Theta_i (t_j)\Phi s_i^T ,
\end{align*}
where $\Theta_i (t_j)$ is a $T_i \times M$ spline matrix on the grid $t_j$, $\Phi$ is the $M \times K$ "loading matrix" with columns $\phi_k$ as spline coefficient vectors, and $s_i^T$ is the vector of scores for subject $i$. The model parameters are estimated using an alternating minimization method (Udell et al., 2016).  The score vector $s_{i}$ can be estimated using a Poisson generalized linear model with identity link and a nonnegative constraint, with $Y_i (t_j)$ being the functional responses and $\Theta (t_j) \Phi$ being the covariate matrix. When estimating the spline coefficient matrix $\Phi$, a second derivative penalty is applied to ensure smoothness. Implementation of NARFD in R is available on GitHub (Backenroth et al., 2020). \\
\\
Similar to PFPCA, NARFD assumes a Poisson data generative process which is well defined for count-valued data. Unlike PFPCA, however, NARFD constrains the functional prototypes to be non-negative. Furthermore, NARFD produces highly-intepretable local functional prototypes which are added together in a straightforward way to reconstruct the observed functions. In contrast, FPCA methods produce components that vary across the entire functional domain without contraints, and are added together via complex cancellations and amplifications to reconstruct the observed functions. If fitted well, NARFD results should be more intepretable than PFPCA results when we deal with nonnegative count-valued accelerometry data. 
\subsection{Results}
\label{subsect:feature}
\begin{figure}[h!]
     \centering
        \includegraphics[width=0.75 \linewidth]{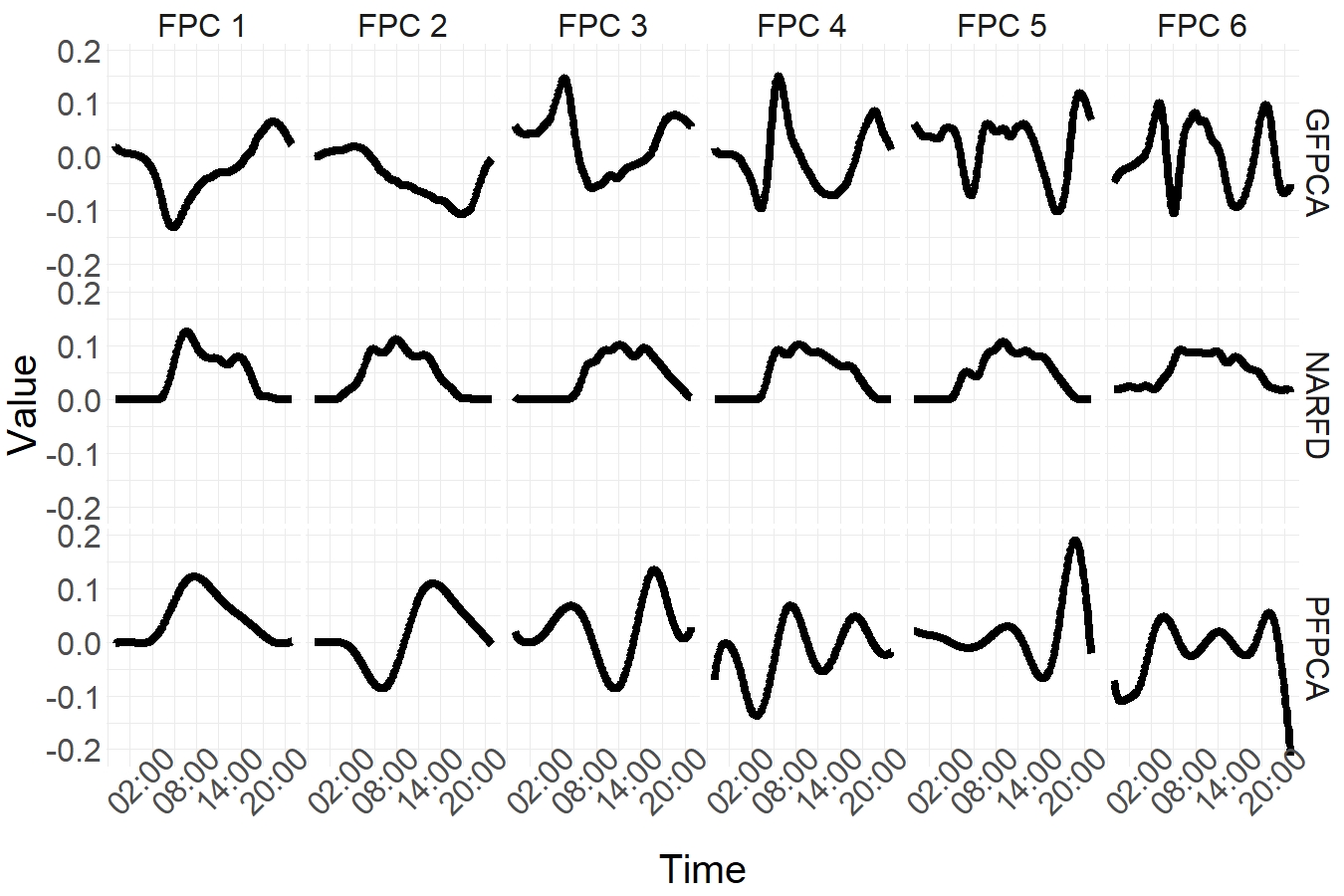}
        \caption{First 6 principal components calculated on the population, 5 minute-level NHANES accelerometry data using the GFPCA, NARFD and PFPCA. }
        \label{fig:2}
    \end{figure}

\begin{figure}[th]
\begin{centering}
   \includegraphics[width=0.75\linewidth]{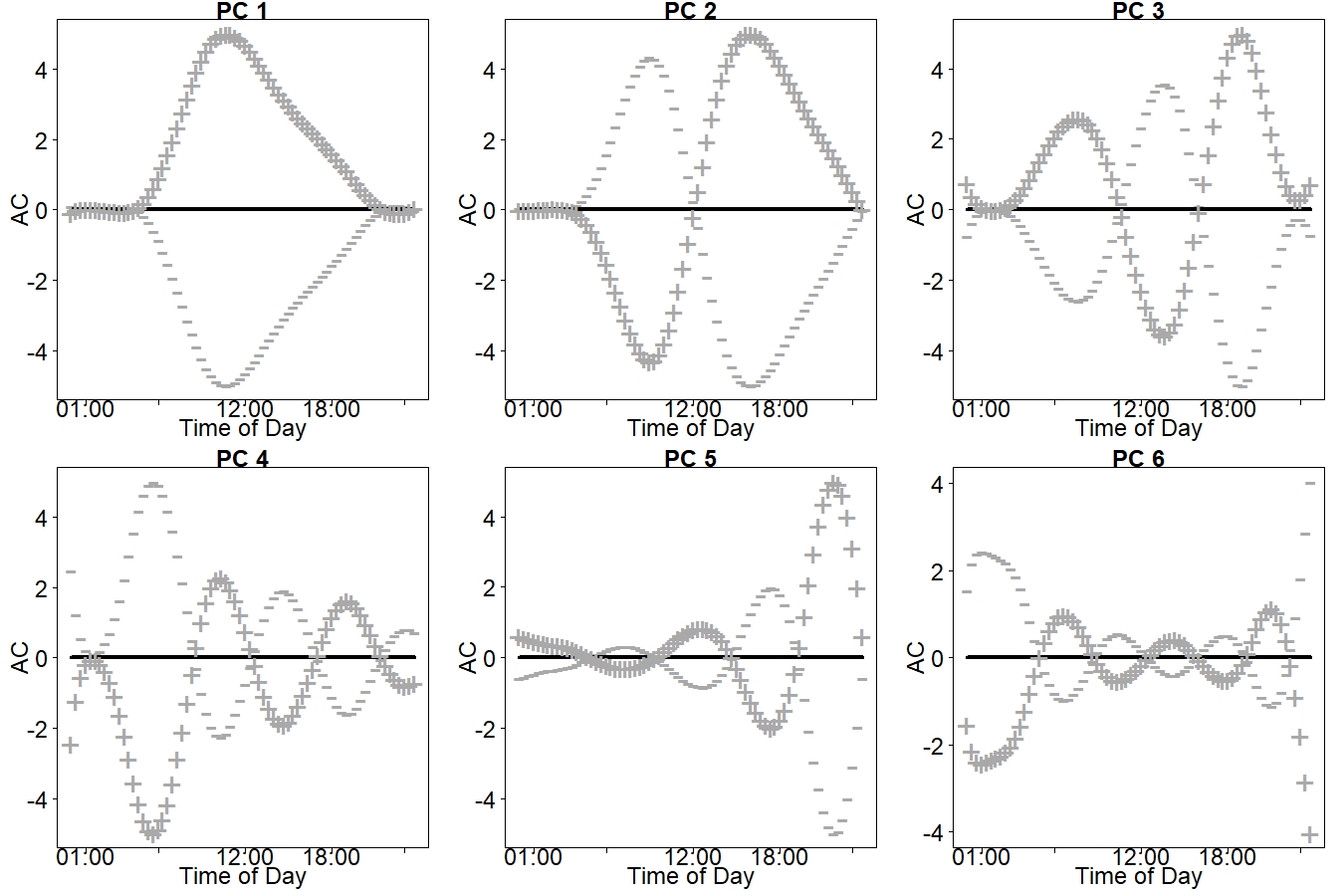}
   \caption[Two numerical solutions]{ First 6 Poisson FPCA principal components calculated on the population, 5 minute-level NHANES accelerometry data using Poisson FPCA. Solid lines represent the base line zero, +, - lines denote the effect of being 2 standard deviation from a score of 0 on the particular principal component. } 
   \label{fig:1}
   \end{centering}
\end{figure}

We extract features that characterize each participant's daily physical activity by fitting GFPCA, PFPCA, and NARFD on the participant's accelerometry record and taking the functional principal component (FPC) scores as the features. The features extracted by FPCA facilitate interpretations of the accelerometry data. For example, a change in one of the FPC scores may indicate certain change of pattern in the participant's daily activity. The FPC scores estimated with the three methods will later be carried forward as predictors in the mortality prediction models. \\
\\
Figure \ref{fig:2} shows principal components for GFPCA, NARFD, and PFPCA. The PFPCA principal components are smooth and the most interpretable. The GFPCA principal components can capture the peaks and valleys at different time points, but some principal components have unwanted non-smoothness. The principal components from the NARFD adopt a local estimation approach, with each one having a peak or valley in a specific area, depending on the score for that component. When applied to accelerometry data, NARFD has been shown to more intepretable than PFPCA (Backenroth et al., 2020). However, the subject-wise median approach to the accelerometry data took away the local estimation advantage from the NARFD functional principal estimation. As a result, each of the six NARFD principal components in the example shows one large, flat plateau during the daytime, which is not helpful for inference. NARFD principal components are unable to characterize short-term activity spikes and drops at different time points. The first four Gaussian FPCs have similar interpretations as the Poisson FPCs, where the change in scores indicates a change in the location of peaks and valleys. However, the fifth and sixth Gaussian FPCs are not smooth and difficult to interpret. This makes Gaussian FPCA a less attractive inference tool than Poisson FPCA.
The conclusion is that Poisson FPC scores characterize the NHANES accelerometry data better than NARFD or Gaussian FPC scores. \\
\\
\\
In Figure \ref{fig:1}, the pluses (and minuses) show what a specific principal component would look like for an individual whose score is two standard deviations above  (below) the mean. For example, in the second principal component from the PFPCA, we can see that a higher score for this component results in an upward shift in the second half of the day and a downward shift in the first half of the day. The interpretation is that a high score in the second principal component denotes an individual with higher than average afternoon/nighttime activity and lower morning activity. For the other Poisson functional principal components, as the score changes, the locations of specific peaks and valleys change. The scores for these components could be interpreted as signifying the degree to which an individual follows the FPC patterns. The same plots for NARFD and GFPCA are included in the Appendix Figure \ref{fig:Appendix1}. \\

\begin{figure}[!th] 
     \centering
        \includegraphics[width=0.75 \linewidth]{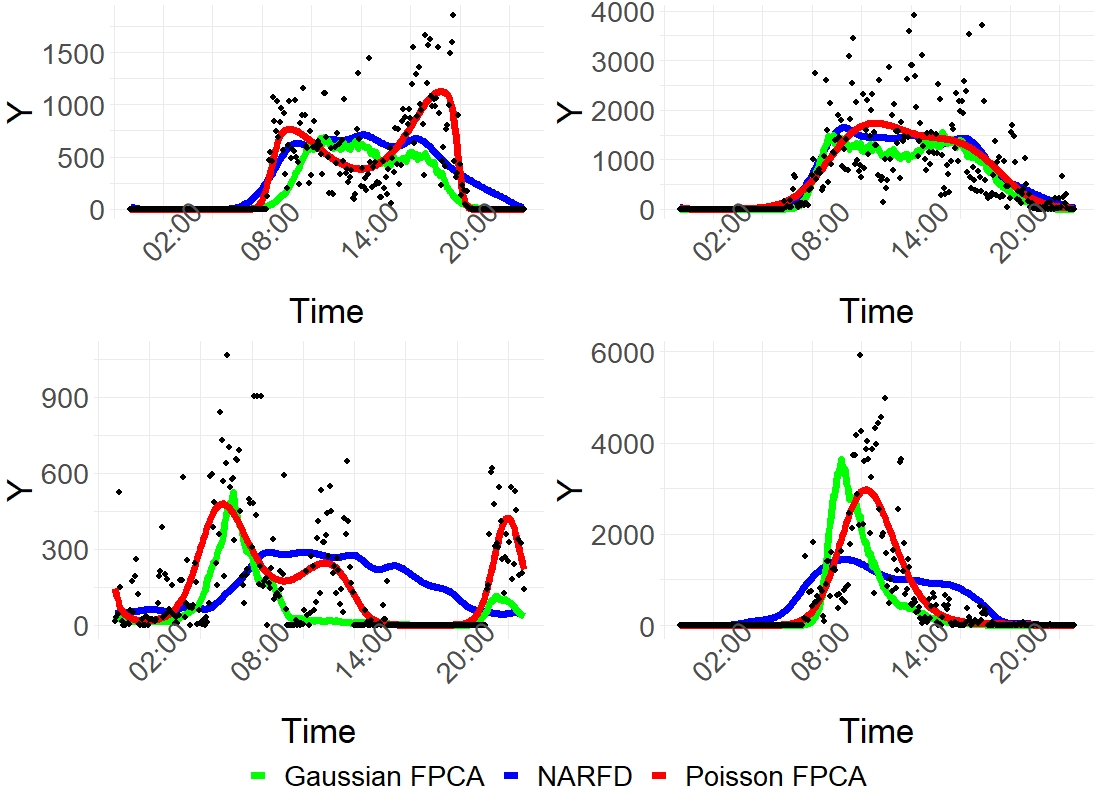}
        \caption{Four randomly-chosen instances of GFPCA,  NARFD, and PFPCA fits}
        \label{fig:3}
    \end{figure}

\begin{figure}[th]
     \centering
        \includegraphics[width=0.8 \linewidth]{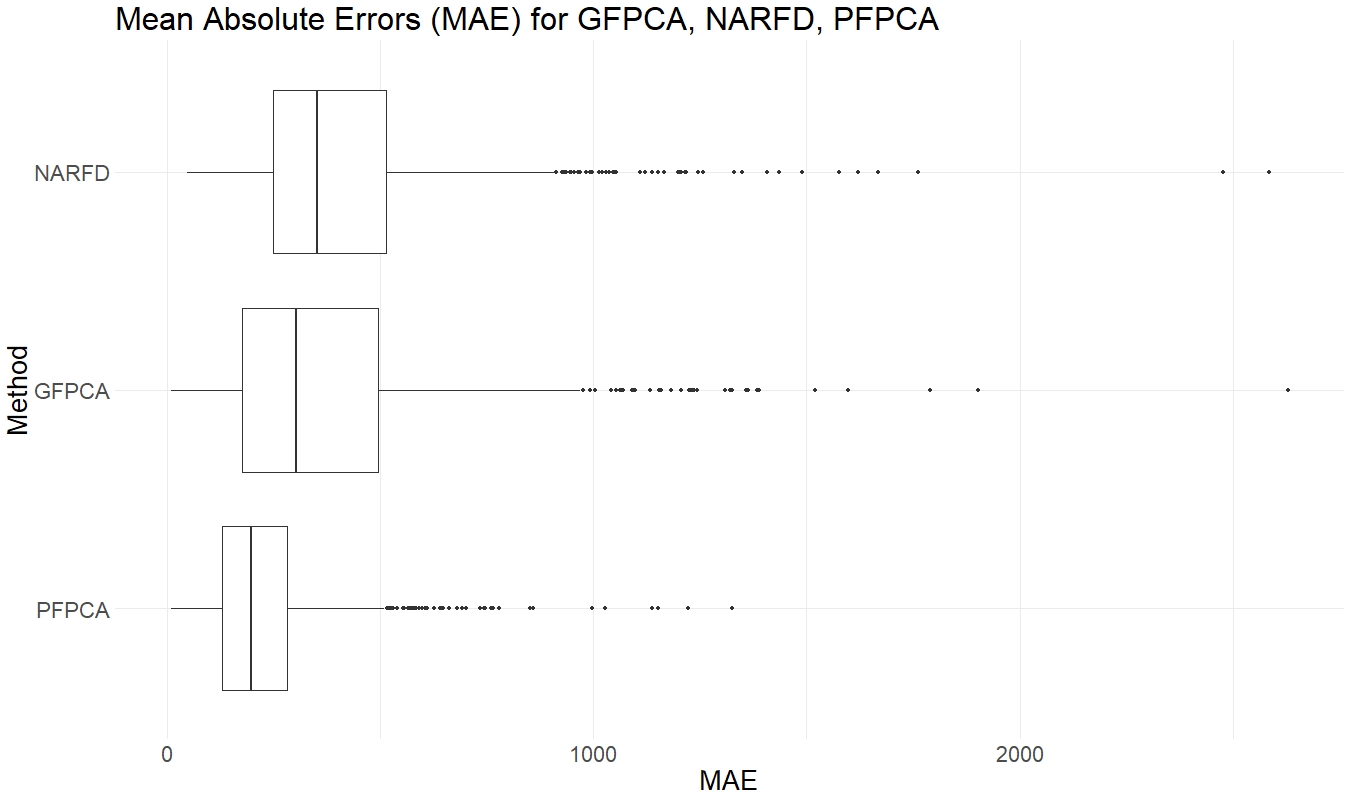}
        \caption{MAE for FPCA models. Poisson FPCA has the best fiA comparison of functional principal component analysis methods with accelerometry applicationsat as it has the smallest MAE. Interestingly,  Poisson FPCA outperforms GFPCA and NARFD models by a wide margin, even though GFPCA and NARFD have relatively similar performance. }
        \label{fig:mae}
\end{figure}

\noindent Figure \ref{fig:3} shows GFPCA, PFPCA, and NARFD fits on four participants' activity curves. The samples are randomly selected. GFPCA fits are overly spiky, even though the fitted values by GFPCA can capture the sudden spikes and dips in the original data better than the PFPCA and NARFD. In contrast, PFPCA produces smooth, reasonable fits that capture the general peaks, valleys, and zeros in the data. NARFD fits are smooth and characterize the nonzero and zero levels in the data, but NARFD fits fail to capture the peaks and valleys of the original data. Overall, PFPCA provides the best fit out of the three models because, unlike NARFD, PFPCA is flexible enough to capture multiple peaks and valleys in the data, and PFPCA fits are less spiky than the GFPCA fits. Figure \ref{fig:mae} depicts the mean absolute errors  (MAE) of the three FPCA models fits for each participant's accelerometry data.  The MAE per participant measures the goodness of fit across three FPCA methods, as methods with lower MAE show superior point estimations relative to other models. PFPCA is shown to be the most accurate model in terms of goodness of fit, because the distribution of PFPCA MAEs has a much lower mean and variance than GPFCA and NARFD and fewer outliers.

\section{Mortality Prediction}
This section replicates the mortality prediction procedure in Leroux's paper { (Leroux et al., 2019)} with the methods in Section \ref{subsect:accel}. We use Gaussian FPCA, Poisson FPCA, and NARFD to extract FPC scores from the accelerometry data and combine the extracted scores with the NHANES covariates to form a set of predictors.  The predictors are then used as the input for the survey-weighted logistic regression, random forest, and AdaBoost models to predict the mortality outcomes. \\
A training/test split is performed on the input data with a split ratio of $70/30$. Given the class imbalance, we perform separate training/test split on the data with "alive" outcomes and the data with "dead" outcomes. The training data with "alive" and "dead" outcomes are combined to form the full training data, likewise for the test data.  After obtaining the prediction results from the nine prediction models, we compare their predictive performance and select the best model.  Finally, we assess how each predictor contributes to the classification of morality outcomes in the best model and tie the interpretations back to the findings in Section \ref{subsect:accel}.  \\
\subsection{Additional Data}
\label{subsect:adata}
In addition to the FPC scores extracted by the three methods in Section \ref{subsect:feature}, we utilize additional data including survey weights, processed covariates, and the five-year all-cause mortality outcomes of the the 2005 - 2006 NHANES cohort {(Leroux et al., 2019)}. NHANES includes a survey weight for each participant to account for the complex NHANES survey design, and it is important to account for the survey weights in our analysis. The processed covariates include demographic, lifestyle and comorbidity information of each participant that we believe are relevant to predicting mortality. Moreover, NHANES data can be linked to US national mortality data to obtain the mortality outcomes of NHANES participants. The mortality data used in the study comes from the R package \textit{rnhanesdata} {(Leroux et al., 2019)}, but the raw mortality data was obtained by linking the 2005 - 2006 NHANES cohort to the death certificate records from the National Death Index (NDI) (National Center for Health Statistics, 2015). Each of these is discussed further below. \\
\subsubsection{Survey Design \& Weights}
\label{subsect:svdesign}
NHANES does not take a simple random sample from the US population; instead it uses a complex survey design. Some of the sampling features include oversampling, adjustment for non-response, and post-stratification. Considering all these design features, NHANES assigns a sample weight to each participant. The survey weight of a participant indicates the number of people in the US population who he/she “represented”, and is the only survey design variable relevant to our study. We need to account for survey weights in our models. Otherwise, our conclusions may be skewed towards the segments of the population that are over-sampled by NHANES, such as racial minorities and the elderly. The raw survey weights come from the R package \textit{rnhanesdata} {(Leroux et al., 2019)}. In \textit{rnhanesdata} package, survey weighting was performed using the ${tableone}$ package (Yoshida and Bartel, (2020)) in R which interfaces with the \textit{survey} package (Lumley, 2020). We utilize the adjusted survey weights in our analysis, defined as the raw survey weights divided by the mean survey weights.  \\
\subsubsection{Covariate Data}
NHANES covariate data contain survey sampling variables (i.e., survey weights) and processed demographic and lifestyle variables for NHANES 2005-2006 participants. The data has 1011 rows and 26 variables with one row per participant in the 2005-2006 wave. The demographic covariates include age, gender, body mass index  (BMI), ethnicity, and education level. The behavioral covariates include smoking status and alcohol consumption. The comorbidities include diabetes, coronary heart disease  (CHD), congestive heart failure  (CHF), cancer and stroke. Other health covariates include blood pressure, high-density lipoproteins (HDL) cholesterol, and total cholesterol. The variables for blood pressure, HDL cholesterol and total cholesterol are directly downloaded from the CDC website, while the remaining variables are from the R package \textit{rnhanesdata} {(Leroux et al., 2019)}. All covariates are included as predictors for mortality prediction. We only include participants that had no missing values for any of the covariates. The processed dataset contains $1003$ subjects after filtering out the $8$ participants with missing mortality outcomes. See Table \ref{tab:1} for more details. 
 
\begin{longtable}[t]{lll}
\caption{Characteristics of individuals included in the analysis}\\
\label{tab:1} \\
\toprule
  & Unweighted & Survey-weighted\\
\midrule
\cellcolor{gray!6}{Sample Size} & \cellcolor{gray!6}{1003} & \cellcolor{gray!6}{1003.00}\\
Number of Weekday (mean (SD)) & 4.55 (0.78) & 4.57 (0.76)\\
\cellcolor{gray!6}{Number of Weekend (mean (SD))} & \cellcolor{gray!6}{1.65 (0.60)} & \cellcolor{gray!6}{1.67 (0.59)}\\
Body Mass Index (\%) &  & \\
\cellcolor{gray!6}{\indent Normal} & \cellcolor{gray!6}{288 (28.7)} & \cellcolor{gray!6}{288.7 (28.8)}\\

\indent Underweight & 11 (1.1) & 12.3 (1.2)\\
\cellcolor{gray!6}{\indent Overweight} & \cellcolor{gray!6}{386 (38.5)} & \cellcolor{gray!6}{380.0 (37.9)}\\
\indent Obese & 318 (31.7) & 322.1 (32.1)\\
\cellcolor{gray!6}{Race (\%)} & \cellcolor{gray!6}{} & \cellcolor{gray!6}{}\\
\indent White & 606 (60.4) & 835.5 (83.3)\\

\cellcolor{gray!6}{\indent Black} & \cellcolor{gray!6}{196 (19.5)} & \cellcolor{gray!6}{75.6 (7.5)}\\
\indent Hispanic & 180 (17.9) & 65.6 (6.5)\\
\cellcolor{gray!6}{\indent Other} & \cellcolor{gray!6}{21 (2.1)} & \cellcolor{gray!6}{26.4 (2.6)}\\
Gender (\% Female) & 461 (46.0) & 518.0 (51.6)\\
\cellcolor{gray!6}{Age (\%)} & \cellcolor{gray!6}{} & \cellcolor{gray!6}{}\\

\indent $[50,60)$ & 372 (37.1) & 524.6 (52.3)\\
\cellcolor{gray!6}{\indent [60,70)} & \cellcolor{gray!6}{346 (34.5)} & \cellcolor{gray!6}{279.6 (27.9)}\\
\indent $[70,80)$ & 210 (20.9) & 158.1 (15.8)\\
\cellcolor{gray!6}{\indent [80,85)} & \cellcolor{gray!6}{75 (7.5)} & \cellcolor{gray!6}{40.6 (4.1)}\\
Education (\%) &  & \\

\cellcolor{gray!6}{\indent Less than High School} & \cellcolor{gray!6}{256 (25.5)} & \cellcolor{gray!6}{144.7 (14.4)}\\
\indent High School & 246 (24.5) & 266.3 (26.6)\\
\cellcolor{gray!6}{\indent College and Above} & \cellcolor{gray!6}{501 (50.0)} & \cellcolor{gray!6}{592.0 (59.0)}\\
Number of Alcoholic Drinks per Week (mean (SD)) & 2.87 (6.13) & 3.16 (6.31)\\
\cellcolor{gray!6}{Cigarette Smoking (\%)} & \cellcolor{gray!6}{} & \cellcolor{gray!6}{}\\

\indent Never & 464 (46.3) & 472.1 (47.1)\\
\cellcolor{gray!6}{\indent Former} & \cellcolor{gray!6}{377 (37.6)} & \cellcolor{gray!6}{371.2 (37.0)}\\
\indent Current & 162 (16.2) & 159.7 (15.9)\\
\cellcolor{gray!6}{Diabetes (\% Yes)} & \cellcolor{gray!6}{127 (12.7)} & \cellcolor{gray!6}{98.9 (9.9)}\\
Congestive Heart Failure (\% Yes) & 27 (2.7) & 22.9 (2.3)\\

\cellcolor{gray!6}{Coronary Heart Disease (\% Yes)} & \cellcolor{gray!6}{57 (5.7)} & \cellcolor{gray!6}{59.6 (5.9)}\\
Cancer (\% Yes) & 134 (13.4) & 137.5 (13.7)\\
\cellcolor{gray!6}{Stroke (\% Yes)} & \cellcolor{gray!6}{24 (2.4)} & \cellcolor{gray!6}{20.9 (2.1)}\\
HDL Cholesterol (mean (SD)) & 56.20 (16.58) & 56.82 (16.08)\\
\cellcolor{gray!6}{Total Cholesterol (mean (SD))} & \cellcolor{gray!6}{204.71 (41.57)} & \cellcolor{gray!6}{206.97 (40.66)}\\

Systolic Blood Pressure (mean (SD)) & 131.14 (20.54) & 129.16 (19.81)\\
\bottomrule
\end{longtable}

\subsubsection{Mortality Data}
\label{subsect:mortdat}
The mortality data we use are from the package \textit{rnhanesdata} {(Leroux et al., 2019)}. While the \textit{rnhanesdata} package provides the processed mortality data, the raw mortality data were obtained by linking the $2011$ death certificate records provided by the National Death Index (NDI) to the NHANES record of 2005 - 2006 participants {(Leroux et al., 2019)}. For each participant, the mortality outcome is $1$ if the participant has died by the year $2011$, and $0$ otherwise. The "alive" participants account for $94 \%$ of all participants, while the remaining $6 \%$ are "dead." For mortality prediction, we treat the mortality outcomes from the mortality data as the response values, and $1003$ subjects that appear in both NHANES covariate and mortality data are included in the analysis.  \\
\subsection{Methods}
For each subject $i$, let $y_i$ be the five-year, all-cause binary mortality outcome. In other words, we define $$y_i = \begin{cases} 1 \text{ if participant i is dead after 5 years, } \\ 0 \text{ if participant i is alive after 5 years. }\end{cases}$$ We use survey-weighted logistic regression, random forest, and AdaBoost methods to predict the mortality outcomes. While each method has been well studied in different aspects, this section introduces the three methods in a way that is most natural and relevant to the mortality prediction problem. \\
\subsubsection{Logistic Regression}
For each subject $i$, we fit a survey-weighted logistic regression model of the following form. 
\begin{equation*}
    logit (y_i|X_i, s_i) = \alpha + X_i^T \beta + s_i^T \theta
\end{equation*}
where $y$ denotes the probability of death in 5 years, $s_i =  (s_{i1}, ..., s_{iK})^T$ is the $K \times 1$ dimensional vectors of FPCA/PFPCA/NARFD scores for subject $i$, and $X_i$ is the predictor matrix that contains the NHANES covariates. The logistic regression is fit using the "svyglm" function from the \textit{survey} package to incorporate the survey weights as observation weights. \\
\subsubsection{Random Forest}
Because logistic regression is based on a linear relationship between predictors and log odds of success, it often fails to capture non-linear effects and interactions between predictors. Random forest is a more suitable prediction model for dealing with non-linear effects. Random forests are based on decision trees. In a decision tree, at each node, one predictor is used to split the tree into two branches. This is done by selecting the predictor and split which minimizes the impurity in the tree. In a random forest model, a large number of decision trees are fit using a random subset of the predictors for each tree. For regression, the predicted value is the average of all the trees values, while for classification, the predicted value is either the most commonly predicted class, or, in the case of two classes represented by 0 and 1, the average predicted value across all trees can be considered a probability that the subject belongs to the class denoted by 1. The added complexity of random forests comes at the cost of interpretability of the model. To predict mortality using our data, we fit a random forest using the "ranger" function from the \textit{ranger} package (Wright and Ziegler, 2017). We chose the \textit{ranger} package over more commonly-used random forest packages such as \textit{RandomForest} because the function "ranger" has the capability of including case weights.
\subsubsection{AdaBoost}
\label{subsect:adaboost}
Random Forest and logistic regression models are known to perform poorly on data with severe class imbalance.  A boosting method is an excellent remedy to the problem. One of the most commonly used boosting methods is AdaBoost. AdaBoost is an iterative classification algorithm that trains decision trees on the misclassified samples from the previous iteration in each iteration. Previous studies have shown that AdaBoost models improve the prediction accuracy of minority samples { (Longadge et al., 2013)}. Therefore, boosting may present an adequate solution to the class imbalance problem in mortality data.  We use an AdaBoost model that assumes an exponential loss on the $0-1$ mortality outcomes, with each observation weighted by the adjusted survey weight. The implementation is done using function "gbm" from the R package \textit{gbm} { (Greenwell et al., 2020)}. 
 \subsection{Results}
 \label{subsect:mort}
\begin{figure}[!th]
     \centering
        \includegraphics[width=0.8 \linewidth]{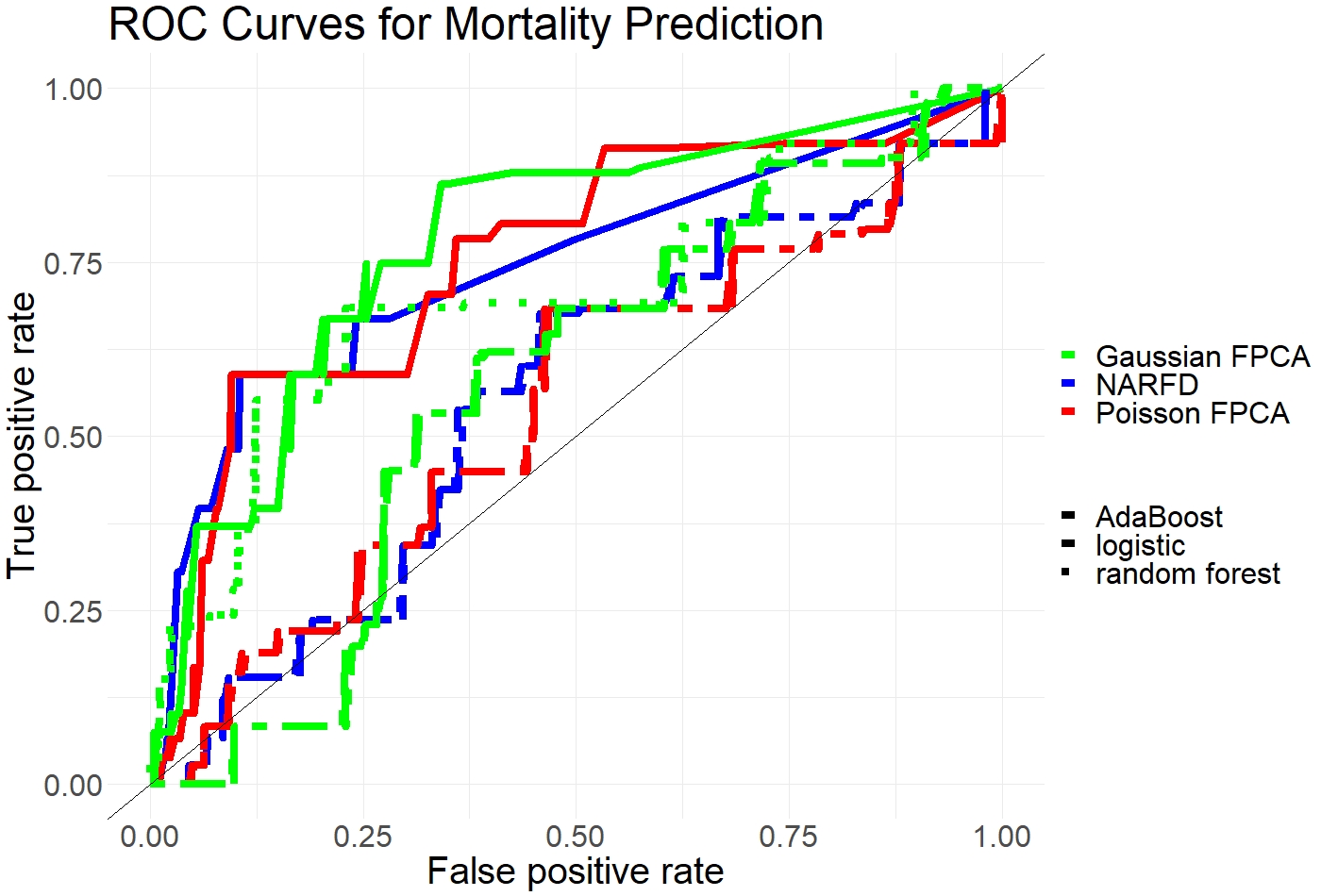}
        \caption{ROC curves of the logistic regression/random forest/AdaBoost with GPFCA/NARFD/PFPCA scores for predicting 5-year all-cause mortality outcomes }
        \label{fig:4}
    \end{figure}

We use logistic regression, random forest, and AdaBoost methods as the main prediction models, and all three models weight observations with the adjusted survey weights. The input data are formed from the NHANES covariates combined with the GFPCA, NARFD, or PFPCA scores. For the logistic regression models, the FPC scores are selected using backward stepwise logistic regression with AIC. For logistic + GFPCA (logistic regression with GFPCA scores) model, score 4 is included. For logistic + NARFD model, no score is included. For logistic + PFPCA model, FPC scores 2, 3, 4, 5, and 6 are included. After obtaining the prediction results, we compare the nine models' performance using survey-weighted ROC curves and AUC scores. We also fit each model type with no FPC scores as a base line. The survey-weighted ROC and AUC are implemented through the function "WeightedROC" from the R package \textit{WeightedROC} { (Hocking, 2020)}. \\
\\
The ROC curves for the prediction models are shown in Figure \ref{fig:4}. For visual clarity, and to focus on the FPCA results, the model with no scores was not included in the ROC curve plot. The AUC scores are shown in Table \ref{tab:2}. From most to least accurate, the models are AdaBoost, logistic regression, and random forest. The accuracy of AdaBoost results from the fact that AdaBoost classifies a portion of "dead" outcomes accurately while logistic regression and random forest fails to predict "dead" outcomes due to class imbalance. Random forest and AdaBoost models outperform logistic regression models. From best to worst, the submodels are GFPCA, PFPCA, NARFD, except for in the logistic regression model. The difference in AUC between GFPCA and PFPCA models is negligible under both AdaBoost and logistic regression. \\
    \begin{table}[th]
    \caption{AUC scores for out-of-sample mortality prediction}
\label{tab:2}
\centering
\begin{tabular}[t]{lllll}
\toprule
Model & GFPCA & NARFD & PFPCA &Baseline (no FPCA scores)\\
\midrule
\cellcolor{gray!6}{AdaBoost} & \cellcolor{gray!6}{0.717} & \cellcolor{gray!6}{0.744} & \cellcolor{gray!6}{0.754}&
\cellcolor{gray!6}{0.729}\\
Logistic & 0.555 & 0.554 & 0.538 &0.555\\
\cellcolor{gray!6}{Random Forest} & \cellcolor{gray!6}{0.724} & \cellcolor{gray!6}{0.703} & \cellcolor{gray!6}{0.722}&
\cellcolor{gray!6}{0.733}\\
\bottomrule
\end{tabular}
\end{table}

\noindent 
\vspace{5mm}\\
We recommend AdaBoost + PFPCA as the go-to method for analyzing future NHANES accelerometry data because PFPCA has better model interpretation than GFPCA and a better AUC. In Section \ref{subsect:feature}, the GFPCA models for the accelerometry data have raised concerns over potential overfitting. On the other hand, PFPCA provides smooth and reasonable fits for each participant's accelerometry data.  Unlike Gaussian models, Poisson models theoretically account for the accelerometry data’s count-valued structure. When it comes to classifying mortality, AdaBoost + PFPCA has the highest AUC scores. \\
\\

\subsubsection{Logistic Regression Inference}
Table \ref{tab:coef1} is the coefficient table of the logistic + PFPCA model. The table contains coefficients and confidence intervals  (CI) for the NHANES covariates and FPC scores selected via AIC; both the coefficients and the confidence intervals are estimated using survey-weighted logistic regression from the function "svyglm." Because our purpose is to infer the effect of each variable on mortality using the coefficient table, the underlying logistic regression is run on the full data rather than the training data. The coefficients table of logistic + NARFD and logistic + GFPCA are included in the Appendix.    \\

\begin{longtable}[t]{ll}
\caption{Coefficients table for logistic + PFPCA \\
*For categorical variable, the base level is given $0$ and $(0,0)$ as the coefficients and $95\%$ confidence intervals.}\\
\label{tab:coef1} \\
\toprule
 Variable & Coefficient (95 \% Confidence Interval) \\
\midrule
\cellcolor{gray!6}{Intercept} & \cellcolor{gray!6}{-6.769 (-16.924, 3.386)}\\
Number of Weekdays & -0.066 (-0.493, 0.361)\\
\cellcolor{gray!6}{Number of weekend} & \cellcolor{gray!6}{0.153 (-0.742, 1.048)}\\
Body Mass Index & -0.013 (-0.097, 0.071)\\
\cellcolor{gray!6}{Race} & \cellcolor{gray!6}{} \\
\indent White & 0 $(0,0)^*$  \\
\cellcolor{gray!6}{\indent Black} &\cellcolor{gray!6}{-0.012 (-0.918, 0.895)}\\
\indent Hispanic & -0.055 (-1.091, 0.981)\\
\cellcolor{gray!6}{\indent Other}  & \cellcolor{gray!6}{-14.02 (-14.987, -13.052)} \\
Gender & \\
\cellcolor{gray!6}{\indent Female}&  \cellcolor{gray!6}{-0.007 (-0.552, 0.537)}\\
\indent Male &   0 $(0,0)^*$  \\
\cellcolor{gray!6}{Age} & \cellcolor{gray!6}{0.107 (0.036, 0.178)} \\
Education  &    \\
\cellcolor{gray!6}{\indent Less than High School} & \cellcolor{gray!6}{$0 (0,0)^*$}  \\
\indent High School & 0.379 (-0.45, 1.208) \\
\cellcolor{gray!6}{\indent College and Above} & \cellcolor{gray!6}{-0.329 (-1.097, 0.44)} \\
Number of Alcoholic Drinks per Week  & 0.034 (-0.018, 0.085) \\
\cellcolor{gray!6}{Cigarette Smoking } & \cellcolor{gray!6}{}\\
\indent Never &   0 $(0,0)^*$  \\
\cellcolor{gray!6}{\indent Former} &\cellcolor{gray!6}{0.548 (-0.585, 1.681)}  \\
\indent Current & 0.74 (-0.086, 1.566)\\
\cellcolor{gray!6}{Has Diabetes} & \cellcolor{gray!6}{0.58 (-0.49, 1.65)} \\
Has Congestive Heart Failure & -0.018 (-1.808, 1.772)\\
\cellcolor{gray!6}{Has Coronary Heart Disease} & \cellcolor{gray!6}{-0.221 (-1.748, 1.306)}\\
Has Cancer & 0.611 (0.208, 1.014) \\
\cellcolor{gray!6}{Has Stroke} & \cellcolor{gray!6}{-0.364 (-2.228, 1.5)}\\

HDL Cholesterol  & -0.013 (-0.031, 0.004)\\
\cellcolor{gray!6}{Total Cholesterol} & \cellcolor{gray!6}-0.002 (-0.011, 0.006){}\\
Systolic Blood Pressure  &  0 (-0.016, 0.017)\\
\cellcolor{gray!6}{$2^{nd}$ score of Poisson FPCA} & \cellcolor{gray!6}{0.006 (-0.002, 0.015)}\\
$3^{rd}$ score of Poisson FPCA & -0.031 (-0.078, 0.016)\\
\cellcolor{gray!6}{$4^{th}$ score of Poisson FPCA} & \cellcolor{gray!6}{-0.015 (-0.046, 0.017)}\\

$5^{th}$ score of Poisson FPCA & 0.019 (-0.016, 0.055)\\
\cellcolor{gray!6}{$6^{th}$ score of Poisson FPCA} & \cellcolor{gray!6}{0.011 (-0.011, 0.032)}\\

\bottomrule
\end{longtable}
\subsubsection{}
\begin{figure}[th]
    \centering
    \includegraphics[width=0.9 \linewidth]{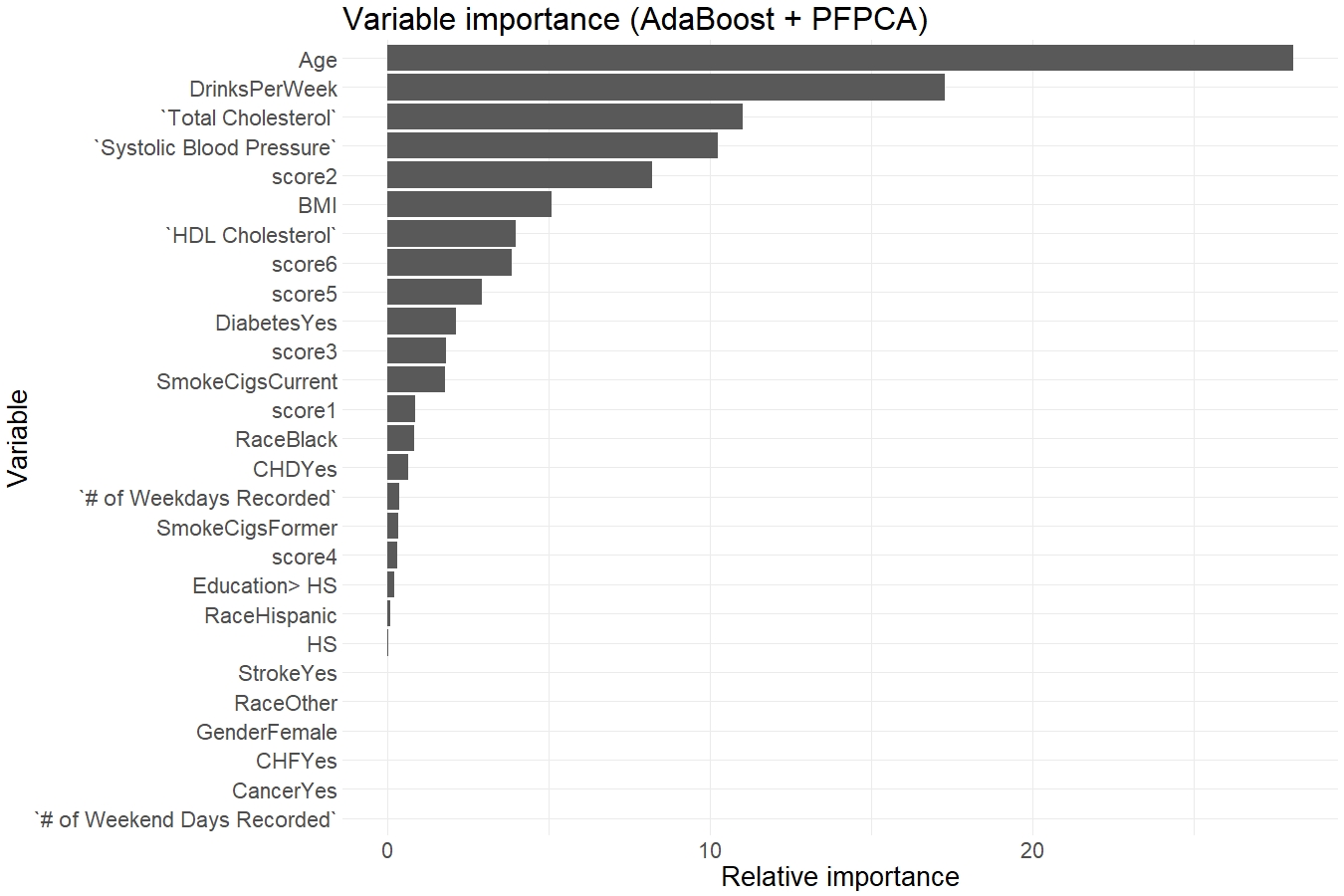}
    \caption{The variable importance plot from AdaBoost + PFPCA. The x-axis is the relative importance. The y-axis is the variable names. Higher relative importance  means larger effect on the mortality outcome. The top five most important variables are age, second PFPCA score, alcoholic drinks per week, blood pressure and  total cholesterol level. In particular, the second PFPCA score is associated with more activity in the late afternoon and less in the morning. }
    \label{fig:5}
\end{figure}

Figure \ref{fig:5} shows the variable importance from the AdaBoost + PFPCA model. Other variable importance plots are included in the Appendix. Among the most important variables are age, the score for the second PFPCA component, alcoholic drinks per week, blood pressure and  total cholesterol level.  Age is to be expected, and is almost always the best predictor of mortality in similar studies. The second PFPCA score is associated with more activity in the late afternoon and less in the morning, as seen in Figure \ref{fig:2}. PFPCA scores 3, 4, 5, and 6 also have nonzero importance, but are much lower in variable importance and, due to their shapes  (Figure \ref{fig:2}),  less interpretable than the second score.
\\
\section{Discussion}
\subsection{FPCA Inference}
The results from Section \ref{subsect:feature} show that PFPCA is the most suitable model for interpreting the NHANES accelerometry data. From looking at the individual FPCs in Figure \ref{fig:2}, the Poisson scores, especially the first two, are easily interpretable. The first score contributes to the overall level of activity, while the second score corresponds to more activity in the afternoon and less in the morning. For the other models, the interpretations are less straightforward. For NARFD, the FPCs all have global effects with slightly different shapes. This was slightly unexpected, as the NARFD model is based off of local estimation. We thought the components would correspond to different local changes, but this was not the case. The Gaussian principal components have some interpretability; for example, the first component seems to be similar to the second Poisson component: more activity in the afternoon and less in the mornings. However, the Gaussian components are more erratic, which may be a indication of over-fitting. This can also be seen in Figure \ref{fig:3}. The Poisson models give a nice middle ground between the overly erratic Gaussian models and the overly smooth NARFD models. Due to the interpretability, the balance between smoothness and fit, and the lowest MAE of the Poisson FPCA, PFPCA is the best candidate of the three FPCA methods for this data.\\

\subsection{Mortality Prediction}
We used logistic regression, random forest, and AdaBoost as the primary mortality prediction models, each with three submodels that include GFPCA, NARFD, or PFPCA scores. The results show that in terms of prediction accuracy, AdaBoost + PFPCA is the best model. AdaBoost and random forest models outperform logistic regression models. We recommend AdaBoost + PFPCA as the go-to model for handling future NHANES mortality data, considering its accuracy and inference power. \\
The most important variables according to the AdaBoost model were age, second component score, number of drinks per week, blood pressure and  total cholesterol level. In the logistic model, age, the $2^{nd}$ Poisson FPC score and the number of drinks per week all had positive coefficients, while total cholesterol level and blood pressure has a near-zero coefficient. This may indicate that age, second component score, and alcohol consumption are all associated with higher risks of mortality for higher values. This lines up with our intuition for age and drinks: given that age and number of drinks have positive coefficients according to the logistic regression, our finding confirms the intuitive idea that the elderly and heavy drinkers have higher mortality risks. \\
The importance of the second score is very interesting, though. Because AdaBoost is based on decision trees, we lose the interpretability of a positive or negative effect of a variable. In the logistic regression, however, the coefficient of the second PFPCA score was positive. It is reasonable to conclude that, because of its positive coefficient in the logistic regression and its high importance in the AdaBoost model, the second PFPCA component is positively related to mortality. That is, people who are more active in the evenings and less so in the mornings are at a higher risk of mortality.\\
\\
Random forest and logistic regression models both suffer from the class imbalance problem. This is due to the fact that our mortality data has a lack of death: $94 \%$ of the mortality outcomes are classified as "alive", while $6 \%$ are classified as "dead". AdaBoost solves the class imbalance problem by retraining the wrong classification outcomes, particularly the false "alive" outcomes. For example, the AdaBoost + PFPCA model was able to predict $2/19$ death outcomes in the test data accurately while the other models failed to produce any accurate death outcomes, which boosted AdaBoost models' predictive accuracy. 

\subsection{Further Work}
There is a multitude of further work that could be done with this data. For example, one could analyze which interactions exist between the variables and include them in the logistic regression, which may lead to a logistic regression model that actually predicts deaths. Another possible direction is to explore the different sub-populations within the data. However, due to the size of this data set and the already existing class imbalance, using a larger data set, or waiting for the ten-year mortality data to ensure enough members of each class would be beneficial, if not necessary, for any sub-population analysis. Additionally, if more data was obtained, it would be useful to perform out of sample predictions using a test data set, to see if the models have generalized predictive power.\\
\\
The methods for analyzing functional count data are not limited to GFPCA, PFPCA, and NARFD. The Poisson distribution has well-known problems such as the lack of flexibility, over- and under-dispersion. Other methods have been proposed to address these problems,  including the STAR framework  ({Kowal and Canale, 2020}) and Zero-inflated Negative Binomial  ({Neelon et al., 2019}). In the future, we hope to design an FPCA method using the STAR framework and compare its inference with Poisson FPCA. \\

\newpage

\section{References}
\begin{enumerate}
    \item Backenroth, D., Shinohara, R., Schrack, J., and Goldsmith, J.  (2020). Nonnegative decomposition of functional count data. \textit{Biometrics}, 76 (4), 1273-1284. \\
    \item Bening, V. E. and Korolev, V. Y.  (2002). Generalized Poisson models and their applications in insurance and finance. \\
    \item Centers for Disease Control and Prevention  (2017) About the national health and nutrition examination survey. \url{http://www.cdc.gov/nchs/nhanes/about_nhanes.htm}
    \item Di, C.Z., Crainiceanu, C.M., Caffo, B.S., Punjabi, N.M.  (2009). Multilevel functional principal component analysis. \textit{Annals of Applied Statistics}, 3 (1):458-488\\ 
    \item Di, J., Leroux, A., Urbanek, J., Varadhan, R., Spira, A. P., Schrack, J., and Zipunnikov, V.  (2017). Patterns of sedentary and active time accumulation are associated with mortality in US adults: The NHANES study. Cold Spring Harbor Laboratory.\\ 
    \item  Gertheiss, J., Goldsmith, J., and Staicu, A.  (2017). A note on modeling sparse exponential-family functional response curves. \textit{Computational Statistics and Data Analysis}, 105, 46-52 \\ 
    \item Greenwell, B., Boehmke, B., Cunningham, J., and GBM Developers  (2020).gbm: Generalized Boosted Regression Models. R package version 2.1.8. \url{https://CRAN.R-project.org/package=gbm}\\
    \item Goldsmith, J., Scheipl, F., Huang, L., Wrobel, J., Di, C., Gellar, J., Harezlak, J., McLean, M., Swihart, B., Xiao, L., Crainiceanu, C. and Reiss, P. (2020). refund: Regression with Functional Data. R package version 0.1-23. \url{https://CRAN.R-project.org/package=refund}
    \item Haslbeck, J., Waldorp, L.  (2020). “mgm: Estimating Time-Varying Mixed Graphical Models in High-Dimensional Data.” \textit{Journal of Statistical Software}, 93 (8), 1–46.\\
    \item Hocking, T.D.  (2020). WeightedR. R package version 2020.1.31  \url{https://github.com/tdhock/WeightedROC}. \\
    \item Kowal, D.  (2019). Integer‐valued functional data analysis for measles forecasting. \textit{Biometrics}, 75 (4), 1321-1333.\\
    \item Kowal, D., Canale, A.  (2020). Simultaneous transformation and rounding  (STAR) models for integer-valued data. \textit{Electronic Journal of Statistics} 14, no. 1, 1744-1772.\\ 
    \item Leroux, A.  (2018). rNHANESdata: NHANES accelerometry data pipeline. R package version 1.0. \url{https://github.com/andrew-leroux/rnhanesdata} \\
    \item Leroux, A., Di, J., Smirnova, E., et al.  (2019). Organizing and analyzing the activity data in NHANES. \textit{Statistics in Biosciences}, 11 (2):262-287. \\ 
    \item Longadge, R., Dongre, S.  (2013). Class Imbalance Problem in Data Mining Review. ArXiv, abs/1305.1707.\\
    \item Lumley, T.  (2010). Complex Surveys: A Guide to Analysis Using R: A Guide to Analysis Using R. John Wiley and Sons.\\
    \item Müller, H.G., Stadtmüller, U.  (2005). Generalized functional linear models. \textit{Annals of Statistics} 33, 774-805.
    \item National Center for Health Statistics (2015) Office of analysis and epidemiology, public-use linked mortality file. \url{http://www.cdc.gov/nchs/data_access/data_linkage/mortality.htm}
    \item Neelon, B.  (2019). Bayesian Zero-Inflated Negative Binomial Regression Based on Pólya-Gamma Mixtures. \textit{Bayesian Anal.} 14, no. 3, 829-855. \\ 
    \item Ramsay, J. O., Silverman, B. W., (1998). Functional Data Analysis / J.O. Ramsay, B.W. Silverman. New York: Springer. Print. \\ 
    \item Shang, H.L.  (2014). A survey of functional principal component analysis. \textit{AStA Adv Stat Anal} 98, 121-142\\ 
    \item Shou, H., Zipunnikov, V., Crainiceanu, C.M., Greven, S.  (2015). Structured functional principal component analysis. \textit{Biometrics.}, 71 (1):247-257.\\  
    \item Simon, N., Friedman, J., Hastie, T., Tibshirani, R.  (2011). “Regularization Paths for Cox's Proportional Hazards Model via Coordinate Descent.” \textit{Journal of Statistical Software}, 39 (5), 1-13.\\
    \item Udell, M., Horn, C., Zadeh, R. and Boyd, S.  (2016). Generalized low rank models. \textit{Foundations and Trends in Machine Learning}, 9, 1-118.\\
    \item  Wright, M.N., Ziegler, A.  (2017). “ranger: A Fast Implementation of Random Forests for High Dimensional Data in C++ and R.” \textit{Journal of Statistical Software}, 77 (1), 1–17.\\
    \item Wrobel, J., Bauer, A., Goldsmith, J., and McDonald, E.  (2020). registr: Curve Registration for Exponential Family Functional Data. R package version 1.0.0. \url{https://CRAN.R-project.org/package=registr}
    \item Yoshida, K., and Bartel, A., (2020). tableone: Create 'Table 1' to Describe Baseline Characteristics with or without Propensity Score Weights. R package version 0.12.0. \url{https://CRAN.R-project.org/package=tableone}
\end{enumerate}
\newpage
\section{Appendix}
\subsection{Further Analysis: Mixed Graphical Model}
 \subsubsection{Method}
 Researchers are sometimes interested in testing the conditional independence between sets of predictors, such as whether race is independent of having diabetes given BMI. The mixed graphical model  (MGM) is particularly suited for the exploratory analysis and hypothesis testing with NHANES data via nodewise regressions for categorical, count-valued, and continuous variables. For a variable $X$, the nodewise regression of variable $X$ is the $L_1$ penalized regression of $X$ on the other variables. The regression assumes Gaussian/Poisson/multinomial distribution for continuous/count/categorical data. The $L_1$ penalized nodewise regression selects the neighborhood of associated variables for each covariate in the data. Previous studies have demonstrated that MGM's power for inferring the relationship between covariates in high-dimensional multivariate data, such as genomics data  (Ying-Wooi Wan et al., 2015).  \\
We use the function "mgm" from the R package \textit{mgm}  (Yang et al., 2015) . The nodewise regression use function "glmnet" from the R package \textit{glmnet} (Friedman et al., 2011). It assigns multinomial, Poisson, Gaussian distributions to fit categorical, count-valued, and continuous-valued variables. 10-fold cross-validation is used to choose the optimal regularization parameter for each regression.\\
The mixed graphical model in the example uses the NHANES covariates with all six PFPCA scores. The graph visualizes the conditional dependency structures between the variables, enabling one to learn between-covariate information that may otherwise be obscured. For example, $6^{th}$ and $4^{th}$ Poisson FPC scores are both positively related to the recorded number of weekends. One possible meaning is that the scores 4 and 6 in PFPCA characterize weekend activities. Furthermore, the number of drinks per week is conditionally independent of age given gender. Given that number of drinks per week and age are the two most important variables for mortality prediction, the graph shows that their effects on mortality are independent given gender. HDL cholesterol and total cholesterol are another pair of interesting predictors. The graph shows that HDL cholesterol and total cholesterol are not closely associated, even though the two variables both measure cholesterol levels. HDL cholesterol positively associates with being female and negatively correlated with BMI. In contrast, total cholesterol negatively associates with having diabetes, coronary heart disease and positively correlated with being "other" races. \\
The graph also provides a fast and intuitive model selection if one wishes to predict other NHANES variables. If one wishes to predict $X_j$ where $1 \leq j \leq n$,  the neighborhood $Nr (X_j)$ of $X_j$ represents the set of covariates that are most indicative of $X_j$.  For example, if $X = Cognitive Heart Failure:Yes$, $Nr (X) = \{Diabetes: Yes, Stroke:Yes, Coronary Heart Failure:Yes, SmokeCigsFormer, Systolic Blood Pressure\}$. It follows that the history of coronary heart failure, diabetes, stroke, whether the person smokes cigarettes and the person's blood pressure are enough to determine whether the person has/will have cognitive heart failure. On the other hand, other covariates such as age, race, gender are independent of having cognitive heart failure given the neighborhood variables.   \\
    \begin{figure}[th]
     \centering
        \includegraphics[width= \linewidth]{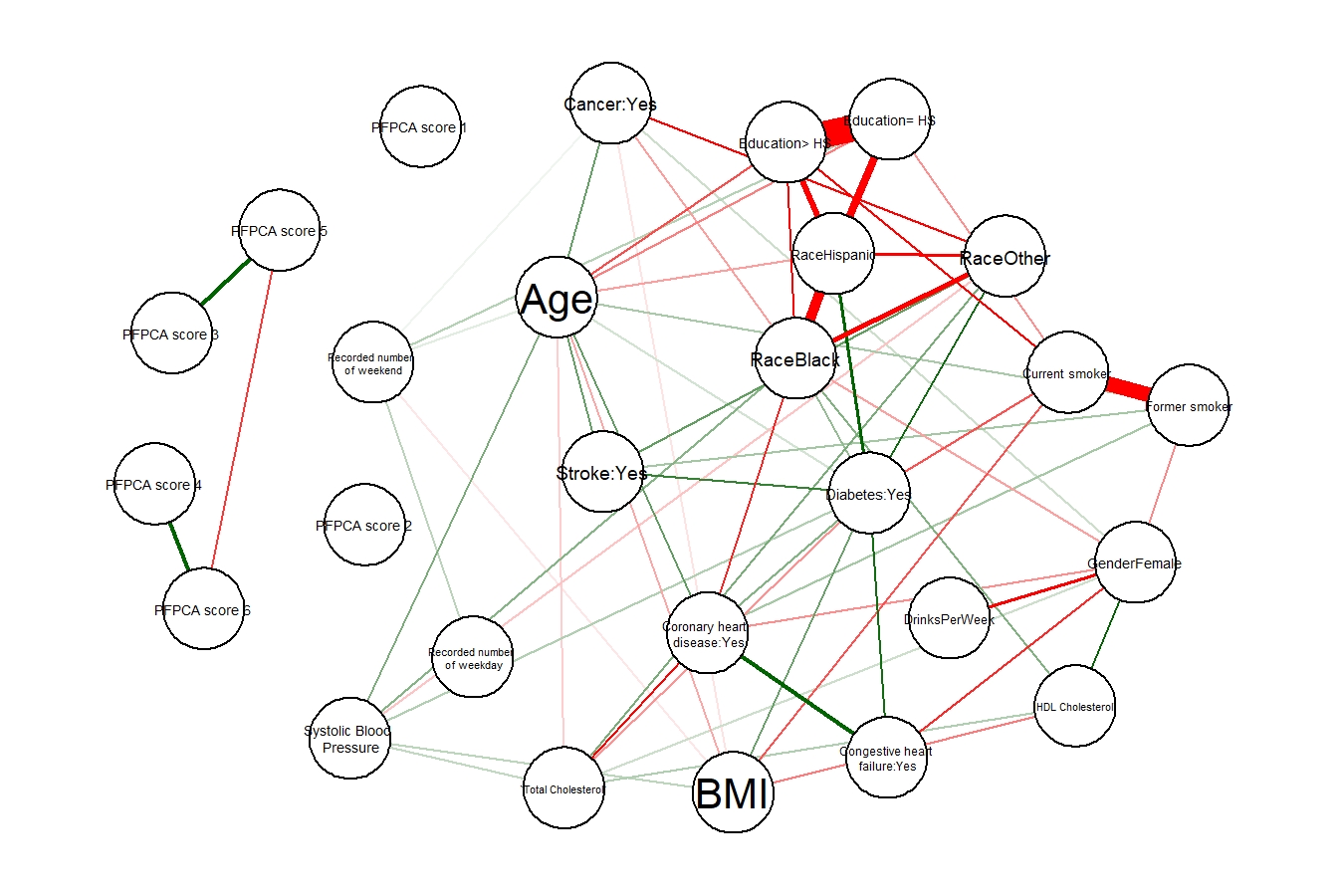} 
        \caption{Mixed Graphical Model for covariates in NHANES. Thicker green edges indicate stronger positive correlation between the two variables; thicker red edges indicate stronger negative correlation between the two variables. }
        \label{fig:6}
    \end{figure}
\newpage
\subsection{Functional principal component plots for GFPCA and NARFD}
\begin{figure} [h!]
\centering
\begin{subfigure}[b]{0.8\textwidth}
   \includegraphics[width=0.9\linewidth]{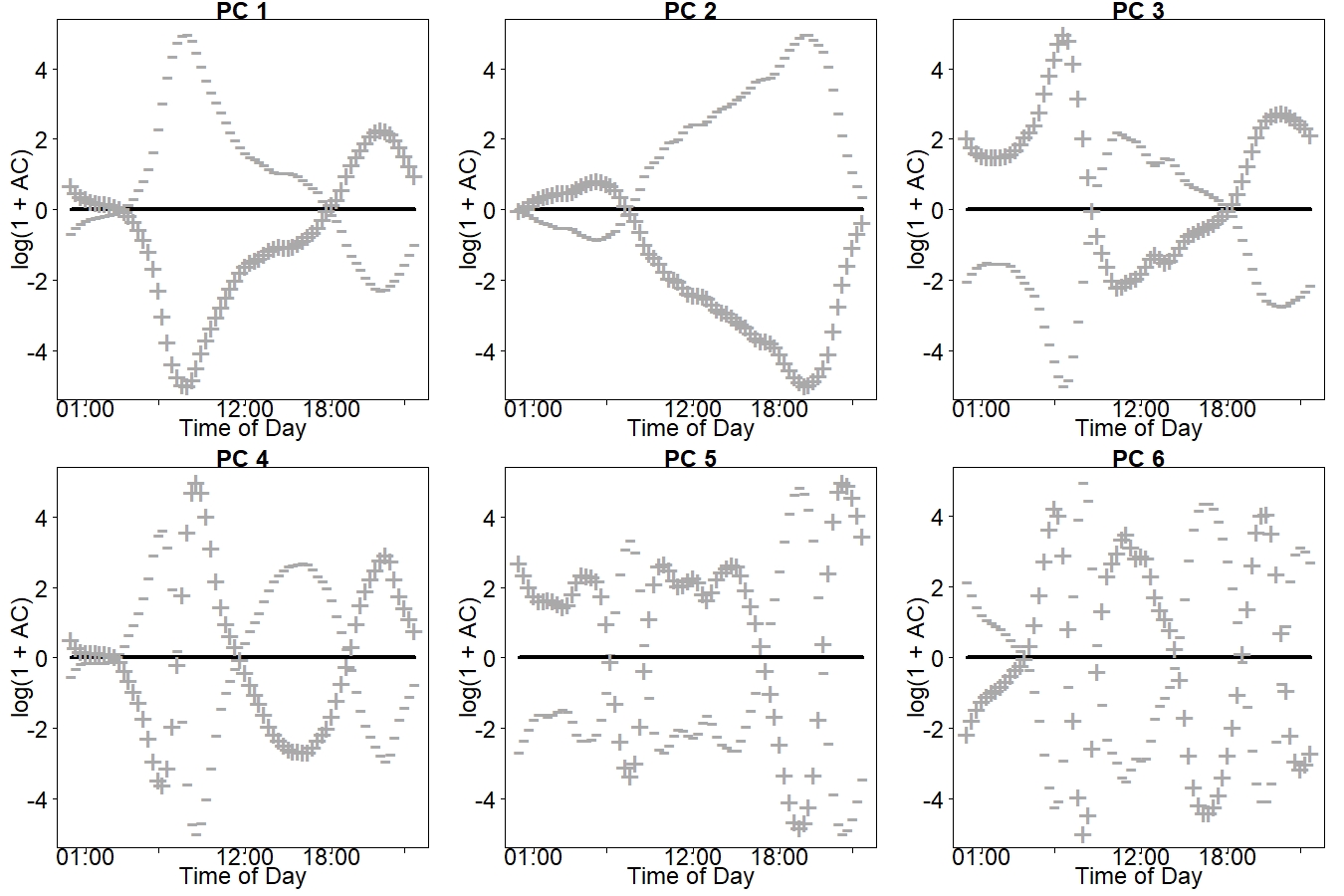}
   \caption{GFPCA}
   \label{fig:Appendix1} 
\end{subfigure}

\begin{subfigure}[b]{0.8\textwidth}
   \includegraphics[width=0.9 \linewidth]{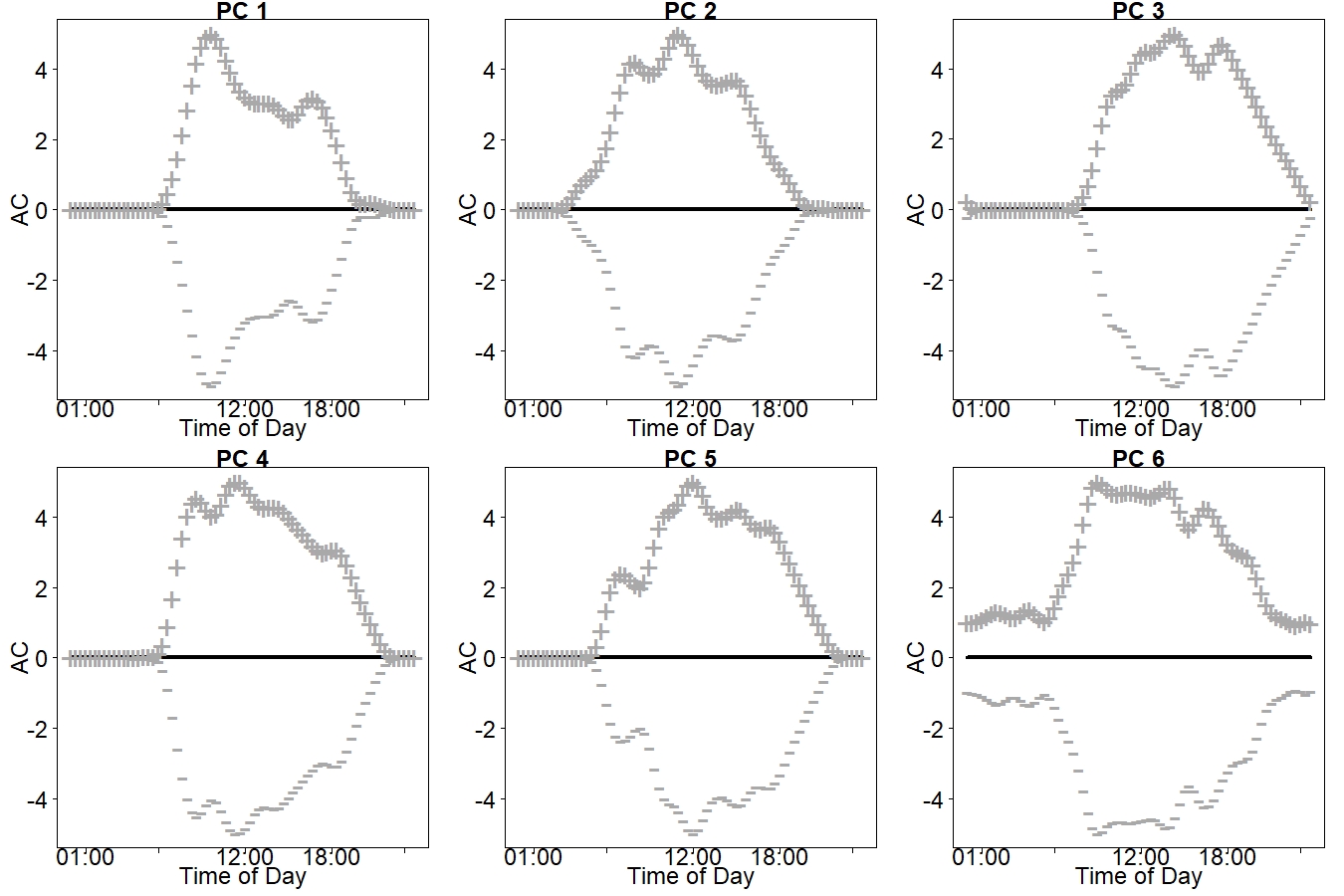}
   \caption{NARFD}
   \label{fig:Ng2}
\end{subfigure}
\caption[Two numerical solutions]{First 6 principal components calculated on the population, 5 minute-level NHANES accelerometry data using GFPCA and NARFD methods. Solid lines represent the base line zero, +,- lines denote the effect of being 2 standard deviation from a score of 0 on the particular principal component. Figure $a)$ for GFPCA. Figure $b)$ for NARFD.} 
   \label{fig:7}
\end{figure}
\subsection{Coefficients Tables for logistic + GFPCA and logistic + NARFD}
\begin{table}[t]

\caption{Coefficents table for logistic + NARFD\\
*For categorical variable, the base level is given $0$ and $(0,0)$ as the coefficients and $95\%$ confidence intervals.}
\centering
\begin{tabular}[t]{ll}
\toprule
Variable & Coefficient(95\% Confidence Interval)\\
\midrule
\cellcolor{gray!6}{Intercept} & \cellcolor{gray!6}{-9.268 (-13.407, -5.129)}\\
Number of Weekdays  & -0.06 (-0.495, 0.374)\\
\cellcolor{gray!6}{Number of weekend} & \cellcolor{gray!6}{0.155 (-0.677, 0.987)}\\
Body Mass Index & -0.007 (-0.092, 0.078)\\
\cellcolor{gray!6}{Race} & \cellcolor{gray!6}{} \\
\indent White & 0 $(0,0)^*$  \\
\cellcolor{gray!6}{\indent Black} &\cellcolor{gray!6}{0.057 (-0.778, 0.893)}\\
\indent Hispanic & -0.009 (-1.027, 1.01)\\
\cellcolor{gray!6}{\indent Other}  & \cellcolor{gray!6}{-13.992 (-14.855, -13.129)}\\
Gender & \\
\cellcolor{gray!6}{\indent Female}&  \cellcolor{gray!6}{0.005 (-0.512, 0.522)}\\
\indent Male &   0 $(0,0)^*$  \\
\cellcolor{gray!6}{Age} & \cellcolor{gray!6}{0.107 (0.036, 0.178)}\\
Education  &    \\
\cellcolor{gray!6}{\indent Less than High School} &\cellcolor{gray!6}{$0 (0,0)^*$}  \\
\indent High School  & 0.394 (-0.404, 1.191)\\
\cellcolor{gray!6}{\indent College and Above}  & \cellcolor{gray!6}{-0.324 (-1.046, 0.399)}\\
Number of Alcoholic Drinks per Week  & 0.037 (-0.019, 0.092)\\
\cellcolor{gray!6}{Cigarette Smoking } & \cellcolor{gray!6}{}\\
\indent Never &   0 $(0,0)^*$  \\
\cellcolor{gray!6}{Former} & \cellcolor{gray!6}{0.558 (-0.586, 1.702)}\\
Current & 0.701 (-0.112, 1.515)\\
\cellcolor{gray!6}{Has Diabetes} & \cellcolor{gray!6}{0.556 (-0.535, 1.648)}\\
\addlinespace
Has Congestive Heart Failure & 0 (-1.838, 1.838)\\
\cellcolor{gray!6}{Has Coronary Heart Disease} & \cellcolor{gray!6}{-0.203 (-1.716, 1.311)}\\
CancerYes & 0.604 (0.182, 1.025)\\
\cellcolor{gray!6}{Has Stroke} & \cellcolor{gray!6}{-0.371 (-2.241, 1.499)}\\
HDL Cholesterol  & -0.013 (-0.031, 0.005)\\
\addlinespace
\cellcolor{gray!6}{Total Cholesterol} & \cellcolor{gray!6}{-0.003 (-0.012, 0.006)}\\
Systolic Blood Pressure & 0 (-0.017, 0.016)\\
\bottomrule
\end{tabular}
\end{table}

\begin{table}[t]

\caption{Coefficents table for logistic + GFPCA \\
*For categorical variable, the base level is given $0$ and $(0,0)$ as the coefficients and $95\%$ confidence intervals.}
\centering
\begin{tabular}[t]{ll}
\toprule
Variable & Coefficient(95\% Confidence Interval)\\
\midrule
\cellcolor{gray!6}{Intercept} & \cellcolor{gray!6}{-8.634 (-12.826, -4.442)}\\
Number of Weekdays & -0.096 (-0.543, 0.351)\\
\cellcolor{gray!6}{Number of weekend} & \cellcolor{gray!6}{0.112 (-0.708, 0.932)}\\
Body Mass Index  & -0.013 (-0.097, 0.071)\\
\cellcolor{gray!6}{Race} & \cellcolor{gray!6}{} \\
\indent White & 0 $(0,0)^*$  \\
\cellcolor{gray!6}{\indent Black} & \cellcolor{gray!6}{0.098 (-0.78, 0.975)}\\
\indent Hispanic & 0.006 (-1.02, 1.033)\\
\cellcolor{gray!6}{\indent Other} & \cellcolor{gray!6}{-14.142 (-15.166, -13.119)}\\
Gender & \\
\cellcolor{gray!6}{\indent Female}& 0.081 (-0.505, 0.666)\\
\indent Male &   0 $(0,0)^*$  \\
\cellcolor{gray!6}{Age} & \cellcolor{gray!6}{0.103 (0.028, 0.177)}\\
Education  &    \\
\cellcolor{gray!6}{\indent Less than High School} &\cellcolor{gray!6}{$0 (0,0)^*$}  \\
\indent High School  & 0.411 (-0.403, 1.226)\\
\cellcolor{gray!6}{\indent College and Above} & \cellcolor{gray!6}{-0.302 (-1.031, 0.427)}\\
Number of Alcoholic Drinks per Week  & 0.036 (-0.021, 0.093)\\
\cellcolor{gray!6}{Cigarette Smoking } & \cellcolor{gray!6}{}\\
\indent Never &   0 $(0,0)^*$  \\
\cellcolor{gray!6}{\indent Former} & \cellcolor{gray!6}{0.553 (-0.571, 1.677)}\\
\indent Current  & 0.782 (-0.091, 1.655)\\
\cellcolor{gray!6}{Has Diabetes}& \cellcolor{gray!6}{0.523 (-0.604, 1.65)}\\
Has Congestive Heart Failure & -0.013 (-1.786, 1.76)\\
\cellcolor{gray!6}{Has Coronary Heart Disease} & \cellcolor{gray!6}{-0.17 (-1.733, 1.392)}\\
Has Cancer  & 0.6 (0.174, 1.026)\\
\cellcolor{gray!6}{Has Stroke} & \cellcolor{gray!6}{-0.5 (-2.342, 1.342)}\\
HDL Cholesterol  & -0.013 (-0.033, 0.006)\\
\cellcolor{gray!6}{Total Cholesterol} & \cellcolor{gray!6}{-0.003 (-0.012, 0.007)}\\
Systolic Blood Pressure & -0.001 (-0.018, 0.015)\\
\cellcolor{gray!6}{$4^{th}$ score of Gaussian FPCA} & \cellcolor{gray!6}{0.057 (0.006, 0.109)}\\
\bottomrule
\end{tabular}
\end{table}

\end{document}